\documentclass[a4paper,,11pt]{article}

\usepackage{ifthen}
\usepackage{ulem}
\usepackage{epsfig}
\usepackage{color}
\usepackage{amssymb}
\usepackage{graphicx}
\usepackage{amsmath}
\usepackage[center]{subfigure}
\usepackage[numbers,sort&compress]{natbib}

\def\be{\begin{eqnarray}}
\def\ed{\end{eqnarray}}

\def\nbr{\bar{n}}

\newcommand{\DS}[1]{/\!\!\!#1}

\makeatletter
\def\fmslash{\@ifnextchar[{\fmsl@sh}{\fmsl@sh[0mu]}}
\def\fmsl@sh[#1]#2{%
  \mathchoice
    {\@fmsl@sh\displaystyle{#1}{#2}}%
    {\@fmsl@sh\textstyle{#1}{#2}}%
    {\@fmsl@sh\scriptstyle{#1}{#2}}%
    {\@fmsl@sh\scriptscriptstyle{#1}{#2}}}
\def\@fmsl@sh#1#2#3{\m@th\ooalign{$\hfil#1\mkern#2/\hfil$\crcr$#1#3$}}
\makeatother

\begin{document}
\begin{flushright}
\end{flushright}
\begin{center}
{\Large\bf  Renormalization-group  analysis of $B \to \pi$ form factors with $B$-meson Light-Cone Sum Rules}\\[2cm]
{\large\bf Yue-Long Shen$^{a}$, Yan-Bing Wei$^{b}$ and Cai-Dian L\"u$^{b}$}\\[0.5cm]
{\it $^a$ College of Information Science and Engineering, Ocean
University of China, Qingdao, Shandong 266100, P.R. China
 }\\
 {\it $^b$Institute of High Energy Physics, CAS, P.O. Box 918(4),
Beijing 100049,   P.R. China}\\[1.5cm]
\end{center}

\begin{abstract}
Within the framework of the $B$-meson light-cone sum rules, we review
the calculation of radiative corrections to the three $B\to \pi$ transition
form factors at leading power in $\Lambda/m_b$. To resum large logarithmic terms, we perform
the complete renormalization-group
evolution of the correlation function. We employ the integral
transformation which diagonalizes evolution equations of the jet
function and the $B$-meson light-cone distribution amplitude
to solve these evolution equations, and obtain renormalization-group improved sum rules for the $B\to \pi$ form factors.
Results of the form factors are extrapolated to the whole physical
$q^2$ region, and are compared with that of other approaches. The
effect of $B$-meson three-particle light-cone distribution
amplitudes, which will contribute to the form factors at
next-to-leading power in $\Lambda/m_b$ at tree level, is not
considered in this paper.

\end{abstract}
\vfill

\section{INTRODUCTION}
The knowledge of the heavy-to-light transition form factors is of
great importance because it is crucial for the determination of 
parameters of the standard model, and for the understanding of strong
interaction dynamics. The $B \to \pi$ transition form factors, which are
closely related to the CKM matrix element $|V_{ub}|$, have been
extensively studied in the literature. Because of the appearance
of the endpoint singularity in the factorization of the
heavy-to-light form factors, the form factors are regarded, in
many approaches, as dominated by long-distance QCD dynamics and
can be calculated only with non-perturbative methods, such as
Lattice QCD, (Light-Cone) QCD Sum Rules, et al. In
\cite{Belyaev:1993wp,Khodjamirian:1997ub,Bagan:1997bp,
Ball:2001fp,Ball:2004ye,Duplancic:2008ix,Bharucha:2012wy,Khodjamirian:2011ub},
the light-cone sum rules (LCSR) with pion light-cone distribution
amplitudes (LCDAs) has been employed to study the $B \to \pi$ form
factors, and next-to-leading-order (NLO) corrections to the
twist-2 and the twist-3 terms as well as renormalization-group (RG)
evolution effects \cite{Kadantseva:1985kb} have been considered. In
this paper, we use the LCSR with $B$-meson LCDAs
\cite{Khodjamirian:2005ea,Khodjamirian:2006st} to calculate the $B
\to \pi$ form factors. The $B$-meson LCSR has also been
established independently in the framework of the soft-collinear
effective theory (SCET) \cite{Bauer:2000yr,Beneke:2002ph},  where
jet functions encoding the ``hard-collinear" dynamics have
been calculated up to ${\cal O}(\alpha_s)$
\cite{DeFazio:2005dx,DeFazio:2007hw}. An alternative approach to
analyse NLO corrections to the sum rules is suggested in
\cite{ymwang:2015wf}, where the ``method of regions"
\cite{Beneke:1997zp} was adopted to compute  the vector form
factor $f_{B \pi}^{+}(q^2)$ and the scalar form factor $f_{B
\pi}^{0}(q^2)$ defined below
\begin{eqnarray}
\langle \pi(p)|  \bar u \gamma^{\mu} b| \bar B (p_B)\rangle &=&
f_{B \pi}^{+}(q^2) \, \left [ p_B + p -\frac{m_B^2-m_{\pi}^2}{q^2}
q  \right ]^{\mu} +  f_{B \pi}^{0}(q^2) \,
\frac{m_B^2-m_{\pi}^2}{q^2} q^{\mu} \,. \label{ffdef}
\end{eqnarray}
Results of the $B$-meson LCSR were shown to be consistent with that of the SCET sum
rules. There is another $B \rightarrow \pi$  transition  form
factor which is defined by the tensor current:
\begin{eqnarray}
\langle\pi(p)|\bar{q}\sigma^{\mu\nu}q_\nu b|B(p_B)\rangle
&=&\frac{if_{B \pi}^T(q^2)}{m_B+m_\pi}[(p_B+p)^\mu
q^2-(m_B^2-m_\pi^2)q^\mu] \,,\label{fTdef}
\end{eqnarray} and this form factor was not computed in \cite{ymwang:2015wf}.
In the heavy-quark limit and at leading order in $\alpha_{s}$, the
three independent form factors are proportional to one universal
form factor at large recoil.  If loop corrections are included,
differences between the three form factors appear. The
hard-spectator-scattering part can be factorized into a
convolution of the perturbative function and the LCDAs of hadrons
($B$ meson and pion) using the QCD factorization approach
\cite{Beneke:2000wa}. Both QCD corrections to the universal form
factor and to symmetry-breaking hard-spectator interactions
have been calculated in \cite{DeFazio:2005dx,DeFazio:2007hw}.
Using the method of regions, QCD corrections to the symmetry-conserving (universal)
form factor and the symmetry-breaking part of the form factors are computed
simultaneously.

The first step towards using
 the method of regions is identifying leading regions which are in
 principle determined by the analytic structure of the Feynman
 diagram \cite{Sterman:1996}. Usually leading regions are closely related to 
 momentum modes of external lines.
In this work, there are three momentum modes from external lines, namely the hard ($b$-quark), the hard-collinear  (interpolating current of pion) and the soft (light-spectator quark) modes. Thus, three momentum regions with scaling behaviors
\begin{eqnarray}
P_{\mu} \equiv (n \cdot P \,, \bar n \cdot P \,, P_{\perp} ) \,, &
\qquad  &
P_{h \,, \,  \mu} \sim {\cal O}(1,1,1) \,,  \nonumber \\
P_{hc \,, \, \mu} \sim {\cal O}(1,\lambda,\lambda^{1/2}) \,, &
\qquad  & P_{s \,, \, \mu} \sim {\cal
O}(\lambda,\lambda,\lambda)\,,
\end{eqnarray}
can contribute to the correlation function at leading power in $\lambda$, where
$\lambda\sim {\Lambda/m_b}$ and $n_{\mu}$ and $\bar n_{\mu}$ are
light-cone vectors, satisfying $n^2=\bar n^2=0$ and $n \cdot \bar
n=2$. In our calculation, only these three regions give leading-power contributions to the correlation function, which is in agreement with the general analysis. The momentum of the fast-moving pion  is chosen to be along
the $\bar n$ direction. This momentum is also chosen to be hard-collinear and
in the Euclidean region ($p^2<0$) to ensure the light-cone
operator-product expansion (OPE) of the correlation function
\cite{Khodjamirian:2005ea,Khodjamirian:2006st}.
The method of regions provides us a natural way to perform
the factorization of the correlation function because
contributions of different momentum regions are considered
individually. It has been shown that the correlation function can
be factorized into the convolution of the hard function, the jet
function and the $B$-meson LCDA which
describe dynamics of the hard, the hard-collinear and the soft
regions, respectively. This procedure is equivalent to the
two-step matching in the SCET where the
hard (jet) function corresponds to the matching coefficient of ${\rm
SCET_I}$ (${\rm SCET_{II}}$) \cite{Beneke:2004rc,Beneke:2005gs}.

The correlation function is factorization-scale independent, thus
the scale dependence of the hard function, the $B$-meson static decay constant, the jet function and
the $B$-meson LCDA must be cancelled, which has been shown at
one-loop level \cite{DeFazio:2007hw,ymwang:2015wf}. At present,
there is still no complete analysis of RG evolutions of all of
the relevant functions. In \cite{ymwang:2015wf}, RG evolutions
of the hard function and the $B$-meson decay constant were performed. The factorization scale in
that paper was chosen to be about $1.5$GeV, which is a typical
hard-collinear scale. This choice is phenomenologically reasonable as the hard-collinear scale is not far
from the non-perturbative scale. RG evolutions of the jet
function and the $B$-meson LCDA are non-trivial since anomalous
dimensions of these two functions are complicated. But on the conceptual side, a
complete RG analysis is necessary.

$B$-meson LCDAs $\phi_B^{-}(\omega)$ and
$\phi_B^{+}(\omega)$ are fundamental inputs of the
$B$-meson LCSR. They are defined by the matrix element
\cite{Grozin:1996pq}
\begin{eqnarray}
&& \langle  0 |\bar d_{\beta} (\tau \, \bar{n}) \, [\tau \bar{n},
0] \, b_{\alpha}(0)| \bar B(p+q)\rangle = - \frac{i \tilde
f_B(\mu) \, m_B}{4}  \bigg \{ \frac{1+ \! \not v}{2} \, \left [ 2
\, \tilde{\phi}_{B}^{+}(\tau) + \left ( \tilde{\phi}_{B}^{-}(\tau)
-\tilde{\phi}_{B}^{+}(\tau)  \right ) \! \not n \right ] \,
\gamma_5 \bigg \}_{\alpha \beta}\,,
\end{eqnarray}
where $[\tau \bar{n}, 0]$ is the Wilson line along the
$\bar{n}$ direction, $v$ and $\tilde{f}_{B}(\mu)$ are the
$B$-meson velocity vector and the $B$-meson static decay constant,
respectively. The Fourier transformation of
$\tilde{\phi}_{B}^{\pm}(\tau)$ leads to
\begin{eqnarray}
\phi_B^{\pm}(\omega)= \int_{-\infty}^{+\infty} \, \frac{d \, \tau}{2 \, \pi} \, e^{i \, \omega \, \tau }   \,
\tilde{\phi}_{B}^{\pm}(\tau-i 0) \,.
\end{eqnarray}
The scale dependence of $\phi_B^{\pm}(\omega)$ has been studied
extensively \cite{Lange:2003ff,Bell:2013tfa}. $\phi_B^{\pm}(\omega)$ obeys the RG
equation with the anomalous dimension being the Lange-Neubert kernel
\cite{Lange:2003ff}. It is difficult to solve the Lange-Neubert
equation in the momentum space. With the eigenfunction of the
Lange-Neubert kernel found in \cite{Bell:2013tfa}, the RG equation
of $\phi_B^{\pm}(\omega)$ can be diagonalized and readily solved.
The RG equation of the jet function can be simplified in the same
way since the scale dependence of the jet function should
be partly cancelled by that of the $B$-meson LCDA.

In this work, we only consider leading-power contributions of
the $B \to \pi$ form factors. $B$-meson three-particle LCDAs
can give subleading-power contributions to the form factors at
leading order (LO) in $\alpha_{s}$ and give leading-power
contributions at NLO \cite{Beneke:2003pa}. Numerically, the
subleading-power correction from three-particle LCDAs is
only a few percent of the leading-power contribution from
two-particle LCDAs  \cite{Khodjamirian:2005ea}. Feynman
diagrams related to three-particle LCDAs can be seen in
\cite{ymwang:2015wf}. The calculation of effects of three-particle LCDAs is expected to be rather complicated, and RG equations
of the three-particle LCDAs are not completely available so far.
Thus we leave this part for the future work.

This paper is arranged as follows. In Section \ref{section: THE
B-to-Pion FORM FACTORS WITH B-MESON LCSRs}  we briefly review the
calculation of the $B \to \pi$ form factors  at NLO and
emphasize symmetry-breaking effects of these form factors.
Then, the RG evolution of the correlation function is shown in details
in the following section. In Section \ref{section: numerical} we
turn to the numerical analysis of the RG improved form
factors. Concluding discussions are presented in Section
\ref{section: summaries}. The Appendix includes two parts, the jet
function in the ``dual" space and dispersion integrals used in
the sum rules.

\section{THE $B \to \pi$ FORM FACTORS WITH B-MESON LCSRs}
\label{section: THE B-to-Pion FORM FACTORS WITH B-MESON LCSRs}

 In order to derive the sum rules for the $B \to \pi$ transition form
 factors, we start with the correlation function
\begin{eqnarray} \Pi^{\mu}(p,q)
&=&  i\int d^4x ~e^{i p\cdot x} \langle 0|T\left\{\bar{d}(x)\DS n
\gamma_5u(x), \bar u(0)\Gamma^\mu
b(0)\right\}|\bar{B}(P_{B})\rangle\,,\label{corre}
\end{eqnarray}
where $\Gamma^\mu=\gamma^\mu(\sigma^{\mu\nu}q_\nu) $ denoting
the vector (tensor) current. According to the Lorentz-structure analysis,
the correlation function is parameterized as $\Pi^{\mu}(n
\cdot p,\bar n \cdot p) =\Pi_n(n \cdot p,\bar n \cdot p) \,  n^\mu
+{\Pi}_{\bar n}(n \cdot p,\bar n \cdot p) \, \bar n^\mu\,$ for
the vector current, and
 $\Pi^{\mu}(p,q) =\Pi_{\rm T}(n\cdot p, \nbr \cdot
p)\epsilon_{\parallel}^{\mu\nu}q_\nu $ for the tensor current, where
the anti-symmetric tensor  $\epsilon_{\parallel}^{\mu\nu}=(n^\mu
\bar{n}^\nu-n^\nu \bar{n}^\mu)/2$. We work in the rest frame of
the $B$ meson and the power-counting rule of the pion
momentum reads
\begin{eqnarray}
n\cdot p\sim {\cal O}(m_b), \,\,\,\,\,\bar{n}\cdot p \sim {\cal
O}(\Lambda).
\end{eqnarray}
In Euclidean region ($\bar{n}\cdot p<0$) the light-cone OPE is
employed to calculate the correlation function. Tree-level
results are written as
\begin{eqnarray}
{\Pi}^{(0)}_{\bar n}(n \cdot p,\bar n \cdot p) &=& \Pi^{(0)}_{\rm
T}(n\cdot p, \nbr \cdot p) =\tilde f_B(\mu) \, m_B \,
\int_0^{\infty} d \omega^{\prime} \,
\frac{\phi_B^-(\omega^{\prime})}{\omega^{\prime} - \bar n \cdot p-
i \, 0}
 \,, \nonumber \\
\Pi^{(0)}_n(n \cdot p,\bar n \cdot p) &=& 0 \,.
 \label{factorization of
correlator:tree}
\end{eqnarray}

The correlation function can also be expressed in terms of the $B
\to \pi$ form factors and the pion decay constant. For
instance
\begin{eqnarray}
\Pi_{\rm T}(p,q) =\frac{i(n\cdot p)^2f_\pi
f^T_{B\pi}(q^2)m_{B}}{(m_\pi^2-p^2)(m_{B}+m_{\pi})}+\int_{\omega_s}^\infty
d\omega \frac{\rho_h(\omega)}{\omega-\nbr \cdot p-i\epsilon},
\end{eqnarray}
where $\rho_h(\omega)$ represents the contribution of excited and continuum states which have the same quantum numbers as pion.
The form factors are extracted by matching the partonic
and the hadronic representations of the correlation function. After
performing the Borel transformation which suppresses the
contribution of excited and continuum states, we obtain the sum
rules for the tensor form factor at LO
\begin{eqnarray}
f^T_{B\pi}(q^2)=\frac{f_B (m_B+m_\pi)}{n\cdot p
f_\pi}e^{m_\pi^2/(n\cdot p \omega_M)}\int^{\omega_s}_0d\omega
e^{-\omega/\omega_M}\phi^B_-(\omega).
\end{eqnarray}
Large-recoil symmetry relations~\cite{Beneke:2000wa} indicate:
\begin{eqnarray}
f^+_{B\pi}(q^2)={m_B\over n\cdot
p}f^0_{B\pi}(q^2)=\frac{m_B}{m_B+m_\pi}f^T_{B\pi}(q^2)=\xi(n\cdot
p),\label{ff relation}
\end{eqnarray}
where $\xi(n\cdot p)$ is the Isgur-Wise function. These symmetry relations
will be broken by QCD corrections.
\begin{figure}[!ht]
\begin{center}
\includegraphics[width=0.7  \columnwidth]{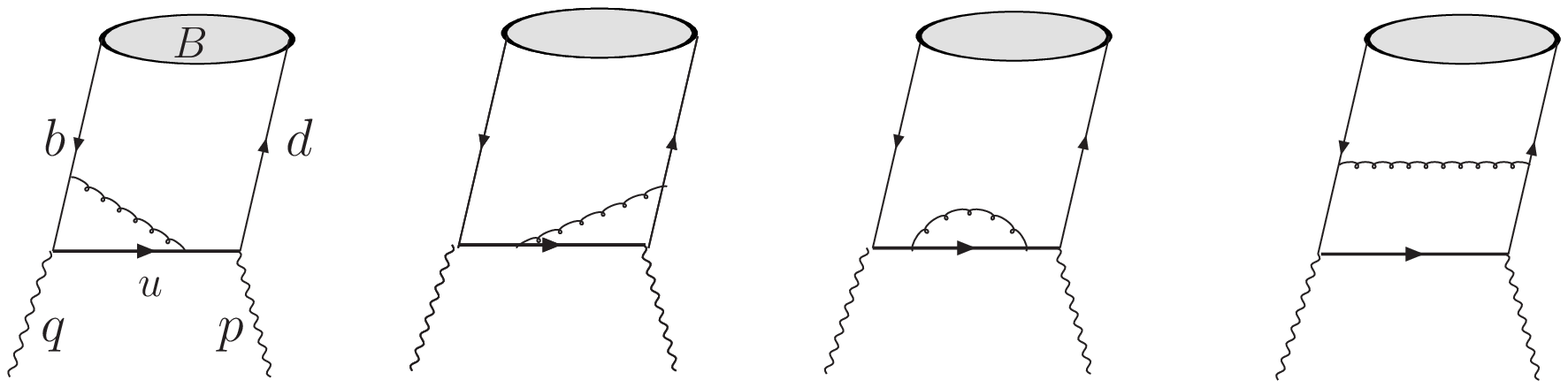}\\
\hspace{0.0 cm}(a) \hspace{2.5 cm} (b)\hspace{2.5 cm} (c) \hspace{2.5 cm} (d) \\
\vspace*{0.1cm} \caption{Diagrammatic representation of the
correlation function $\Pi^{\mu}(n \cdot p,\bar n \cdot p)$ at next-to-leading order in $\alpha_s$. } \label{fig: NLO diagrams of the
correlator}
\end{center}
\end{figure}

\begin{figure}[!ht]
\begin{center}
\includegraphics[width=0.6  \columnwidth]{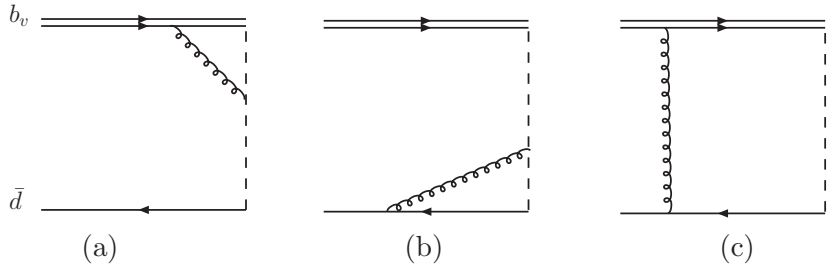}\\
\hspace{0.0 cm}(a) \hspace{3.5 cm} (b)\hspace{3.5 cm} (c)  \\
\vspace*{0.1cm} \caption{One-loop-level diagrams of the $B$-meson
DA $\Phi_{b \bar u}^{\alpha \beta}(\omega^{\prime})$. } \label{fig:
soft subtraction}
\end{center}
\end{figure}

The method of computing radiative corrections to the correlation
function has been introduced in
\cite{ymwang:2015wf}. Adopting the diagrammatic factorization
method \cite{DescotesGenon:2002mw}, the
 hard-scattering kernel at NLO is  determined
by the matching condition
\begin{eqnarray}
\Phi_{b \bar d}^{(0)} \otimes  T^{(1)} = \Pi_{b \bar
d}^{(1)} -  \Phi_{b \bar d}^{(1)} \otimes  T^{(0)} \,,
\label{matching condition of T1}
\end{eqnarray}
where the first and the second terms on the right-hand side of the equation correspond
to full-theory diagrams (Fig.\ref{fig: NLO diagrams of the
correlator}) and effective diagrams (Fig.\ref{fig: soft
subtraction}), respectively. The Lorentz index ``$\mu$" is suppressed in this equation. The definition of $\Phi_{b \bar d}^{(0,1)}$ can be seen in
\cite{ymwang:2015wf}. It has been proved that soft dynamics are
completely cancelled between $\Pi_{b \bar
d}^{(1)}$ and
$\Phi_{b \bar d}^{(1)} \otimes  T^{(0)}$. Thus the hard-scattering
kernel $T$  contains only contributions from hard and
hard-collinear regions at leading power in $\lambda$, with the hard-region and the hard-collinear-region contributions corresponding to
the hard function and the jet function,
respectively. The correlation function is factorized as:
\begin{eqnarray}
\Pi_a(p,q) &=& \tilde{f}_B(\mu) \, m_B \sum \limits_{k=\pm} \,
C_a^{(k)}(n \cdot p, \mu) \, \int_0^{\infty} {d \omega \over
\omega- \bar n \cdot p}~ J_a^{(k)}\left({\mu^2 \over n \cdot p \,
\omega},{\omega \over \bar n \cdot p}\right) \,
\phi_B^{k}(\omega,\mu), \label{factorization of correlator:tree} \,
\end{eqnarray}
where $a=n, \, {\bar n}, \, T$. Results of $C_n, \, C_{\bar n}, \, J_n,
\, J_{\bar n}$ have been given in \cite{ymwang:2015wf}. $C_T$ and
$J_T$ can be calculated following the same method. Nevertheless,
using symmetry relations between the form factors, we are able to obtain
$C_T$ and $J_T$ without repeating the calculation. Symmetry
relations in Eq.(\ref{ff relation}) are broken by loop corrections.
The difference between $C_T$ ($J_T$) and $C_{\bar n}$ ($J_{\bar n}$)
comes from the symmetry-breaking effect. At one-loop level
large-recoil relations can be written as \cite{Beneke:2000wa}:
\begin{eqnarray}
f^0_{B\pi} &=& f^+_{B\pi}[1+{\alpha_s\over
4\pi}(2+{2r\over \bar r}\ln r)]r+{\alpha_s\over 4\pi}\Delta f_0,
\nonumber \\
f^T_{B\pi} &=& f^+_{B\pi}[1+{\alpha_s\over
4\pi}(2\ln{m_b^2\over \mu^2}-{2r\over \bar r}\ln
r)]r+{\alpha_s\over 4\pi}\Delta f_T, \label{ffbreak}
\end{eqnarray}
where $r=n\cdot p/m_{b}$. The first line of above
equations has been confirmed in
\cite{DeFazio:2005dx,ymwang:2015wf}.

The first kind of symmetry-breaking effects, which is shown in
the square brackets of Eq.(\ref{ffbreak}), arises from the hard
function of weak-vertex correction Fig.\ref{fig: NLO diagrams of
the correlator}(a). The weak tensor current $\bar
u(0)i\sigma^{\mu\nu}q_\nu b(0)$  is not a conserved current, thus
there exists operator-renormalization contribution to the
correction function of the tensor current. This contribution
produces an additional term to the hard function
\begin{eqnarray}
\label{tensor.current.syv}
C^{(-)}_{T}(n\cdot p,\mu,\nu)=C^{(-)}_{T}(n\cdot p,\mu)+\delta
C^{(-)}_{T}(n\cdot p,\mu,\nu),
\end{eqnarray}
where $\nu$ is the renomalization scale. $\delta C^{(-)}_{T}(n\cdot
p,\mu,\nu)$ corresponds to the $\ln {m_b^2\over \mu^2}$
term ($\mu$ should be changed to $\nu$) in Eq.(\ref{ffbreak}).
Inserting the hard function of $f^+_{B\pi}$ \cite{ymwang:2015wf} into Eq.(\ref{tensor.current.syv}),
we  obtain
\begin{eqnarray}
C^{(-)}_T(n\cdot p, \mu,\nu) &=& 1 - \frac{\alpha_s \, C_F}{4 \,
\pi}\,  \bigg [ 2\ln\frac{\nu}{m_b}+2\ln^2 {\mu \over m_b}
-(4\ln r-5)\ln {\mu \over m_b}  \nonumber \\
&&+2\ln^2 r  +  2{\rm Li_2} \left (\bar r  \right )  
- {4r-2\over r-1} \, \ln r+{\pi^2 \over 12} + 6 \bigg ].
\label{results of hard coefficients2}
\end{eqnarray}
 The second kind of symmetry-breaking effects
$\Delta f_T$ which corresponds to Fig.\ref{fig: NLO diagrams of
the correlator}(b,c), comes from the ``hard spectator"
contribution in the QCD factorization.  Since only the hard-collinear
region contributes to Fig.\ref{fig: NLO diagrams of the
correlator}(b,c) at leading power in $\lambda$, $\Delta f_T$ is
related to jet functions. The jet function $J_T^{(-)}$, which equals
$J_{\bar n}^{(-)}$, is the symmetry-conserving term. Only
$J_T^{(+)}$ corresponds to the symmetry-breaking effect.
Comparing $\Delta f_T$ with $\Delta f_0$, and employing the result
of $J^{(+)}_{n,\bar n}$, we find
 \begin{eqnarray}
J_T^{(+)}&=&  -\frac{\alpha_s \, C_F}{4 \,
\pi}(1+r)(1+\frac{1}{\eta}) \ln(1+\eta) \label{results of jet
functions},
\end{eqnarray}
where $\eta=-\omega/\bar n\cdot p$.

\section{RG evolution}

The hard and the jet functions contain  logarithmic terms
such as $\ln^2\frac{\mu}{n\cdot p}$ and $\ln\frac{\mu}{n\cdot p}$,
which become large when the factorization scale $\mu$ is
much smaller than $n \cdot p$. The RG equation approach can be
used to resum large logarithms to all orders in $\alpha_{s}$.
In the following, we will present details of the RG evolution of
$\Pi_{\rm T}(n\cdot p)$, and the approach can be easily
generalized to $\Pi_n(n\cdot p)$ and $\Pi_{\bar n}(n\cdot p)$.

Eq.(\ref{results of hard coefficients2}) contains the operator
renormalization of the tensor current, so we need to perform
evolutions both to the renormalization and the factorization
scales. RG equations governing the renormalization-scale and
the factorization-scale dependence are given by
\begin{eqnarray}{d \over d \ln \nu} {C}^{(-)}_T(n \cdot p, \mu,\nu)&=&
\gamma_T(\alpha_s)  {C}^{(-)}_T(n \cdot p, \mu,\nu) \,, \nonumber \\
{d \over d \ln \mu}{C}_T^{(-)}(n \cdot p, \mu,\nu)&=&
\Gamma_C(\alpha_{s}) {C}^{(-)}_{T}(n \cdot p, \mu, \nu) \,, \label{general RGE of
tildeC1}
\end{eqnarray}
where
\begin{eqnarray}
\Gamma_{C}(\alpha_{s})=-  \Gamma_{\rm cusp}(\alpha_s) \ln { \mu \over n
\cdot p} + \gamma_h(\alpha_s).
\end{eqnarray}
Solutions to Eq.(\ref{general RGE of tildeC1}) are written by
\begin{eqnarray}
 {C}^{(-)}_T(n \cdot p, \mu,\nu)&=&
\exp\bigg[\int^{\alpha_s(\nu)}_{\alpha_s(m_b)}d\alpha_s\frac{\gamma_{T}(\alpha_s)}{\beta(\alpha_s)}\bigg
] {C}^{(-)}_T(n \cdot p, \mu,m_b)\,
\nonumber \\
{C}_T^{(-)}(n \cdot p, \mu,m_b) &=&
\exp\bigg\{\int^{\alpha_s(\mu)}_{\alpha_s(\mu_{h})}d\alpha_s\bigg[\frac{\gamma_{h}(\alpha_s)}{\beta(\alpha_s)}+
\frac{\Gamma_{\rm
cusp}(\alpha_s)}{\beta(\alpha_s)}\bigg(\ln{n\cdot p\over
\mu_h}-\int^{\alpha_s}_{\alpha_s(\mu_h)}\frac{d\alpha'_s}{\beta(\alpha'_s)}\bigg)\bigg]\bigg
\} \, \nonumber \\ &\times &{C}_T^{(-)}(n \cdot p, \mu_{h},m_b)
\,\label{evofun}.
\end{eqnarray}
As the hard function has been calculated at one-loop level,
evolution functions of the hard function are also required to be expanded to ${\cal
O}(\alpha_s)$. The beta function appearing in  Eq.({\ref{evofun}})
reads
\begin{eqnarray}
\beta(\alpha_s) &=& -2\alpha_s\sum_{n=0}\beta_n{ \,({\alpha_s
\over 4 \pi})^{n+1}} \,,\label{tlhard}
\end{eqnarray}
 thus anomalous dimensions $\gamma_h$ and  $\gamma_T$ need to be expanded to
two-loop level:
\begin{eqnarray}
\gamma_h(\alpha_s)&=& {\alpha_s \, C_F \over 4 \pi} \, \left [
\gamma_h^{(0)}
+ \left ({\alpha_s \over 4 \pi} \right )\, \gamma_h^{(1)} + ... \right ] \,, \nonumber \\
\gamma_{T}(\alpha_s) &=& {\alpha_s \, C_F \over 4 \pi} \, \left
[-2 + {\alpha_s \over 4 \pi} \left(19C_F-\frac{257}{9}C_A
+\frac{52}{9}n^{\prime}_fT_F\right)+ ... \right ],\end{eqnarray} 
where $n^{\prime}_{f}$ is the number of the active quark flavors. While
the cusp anomalous dimension $\Gamma_{\rm cusp}(\alpha_s)$ should
be expanded  to
 three-loop level:
\begin{eqnarray}
\Gamma_{\rm cusp}(\alpha_s) &=& {\alpha_s \, C_F \over 4 \pi} \,
\left [ \Gamma_{\rm cusp}^{(0)} + \left ({\alpha_s \over 4 \pi}
\right )\, \Gamma_{\rm cusp}^{(1)} + \left ({\alpha_s \over 4 \pi}
\right )^2 \, \Gamma_{\rm cusp}^{(2)} + ... \right ]
\,,\label{tlhard}
\end{eqnarray}
since the factor $\int^{\alpha_s}_{\alpha_s(\mu_h)}d\alpha'_s/\beta(\alpha'_s)$ starts at $\mathcal{O}(\alpha^{-1}_{s})$.

Note  that the $\nu$ dependence of the form factor must be
cancelled by that of the Wilson coefficient of tensor current. For
phenomenological applications, this renormalization scale can be fixed at $\nu=m_b$. Then
the corresponding evolution kernel reduces to 1, and the Wilson coefficient should be evolved to $m_b$. We rewrite the
second equation of Eq.({\ref{evofun}}) as
\begin{eqnarray}
{C}_T^{(-)}(n \cdot p, \mu) &=& U_1(n \cdot p,\mu_{h},\mu ) \,
{C}_T^{(-)}(n \cdot p, \mu_{h}) \,,
\end{eqnarray}
where the specific expression of $U_1(n \cdot p,\mu_{h},\mu )$ at
${\cal O}(\alpha_s)$ can be found in the appendix of
\cite{Beneke:2011nf}.

RG equations of the jet function and
the $B$-meson LCDA have following forms
\begin{eqnarray}
&& {d \over d \ln \mu} {J}_T^{(-)}\left({\mu^2 \over n \cdot p \,
\omega},{\omega \over \bar n \cdot p}\right) = \left [ \Gamma_{\rm
cusp}(\alpha_s) \ln { \mu^2 \over n \cdot p\, \omega} \right ]
{J}_T^{(-)}\left({\mu^2 \over n \cdot p \, \omega},{\omega \over \bar n \cdot p}\right) \nonumber \\
&& \hspace{3 cm} +   \, \int_0^{\infty} \, d \omega^{\prime}  \,
\omega \,\, \Gamma(\omega,\omega^{\prime},\mu) \,\,
{J}_T^{(-)} \left({\mu^2 \over n \cdot p \, \omega^{\prime}},{\omega^{\prime} \over \bar n \cdot p}\right)  \,,  \\
&& {d \over d \ln \mu} \phi_B^-(\omega, \mu)  =  - \left
[\Gamma_{\rm cusp}(\alpha_s) \ln { \mu \over \omega}
+\gamma_+(\alpha_s)\right ]
 \phi_B^-(\omega, \mu)   \nonumber \\
&& \hspace{3.5 cm} -  \int_0^{\infty} \, d \omega^{\prime}  \,
\omega \,\, \Gamma(\omega,\omega^{\prime},\mu) \,\,
 \phi_B^-(\omega^{\prime}, \mu) \,.
\end{eqnarray}
These evolution kernels are non-diagonal. It is thus difficult to
solve above RG equations in the  momentum space. An
alternative method was suggested in  \cite{Bell:2013tfa}, where
the LN kernel can be diagonalized by translating to the ``dual"
space. It was found in \cite{Braun:2014owa} that the basis of the
dual space is the eigenfunction of the generator of special
conformal transformations. The specific form of the transformation
can be written by:
\begin{eqnarray}
\rho_B^-(\omega',\mu)=\int^\infty_0\frac{d\omega}{\omega'}J_0(2\sqrt{\frac{\omega}{\omega'}})\phi_B^-(\omega,\mu).
\end{eqnarray}
The dual-space LCDA $\rho_B^-(\omega',\mu)$ satisfies a simpler
RG equation
\begin{eqnarray}
{d \over d \ln \mu} \rho_B^-(\omega', \mu)  = \Gamma_\rho(\mu)
\rho_B^-(\omega', \mu)\,,
\end{eqnarray}
where $\Gamma_\rho(\mu)=-\Gamma_{\rm cusp}(\alpha_s) \ln { \mu
\over \hat{\omega}'} -\gamma_+(\alpha_s)$, and
$\hat{\omega}'=e^{-2\gamma_E}{\omega}'$.
 Solving the RG equation
of $\rho^{-}_{B}(\mu)$, one has
\begin{eqnarray}
&&  \rho_B^-(\omega', \mu)  =
e^{V(\mu,\mu_0)}\left(\frac{\mu_0}{\hat{\omega}'}\right)^{-g(\mu,\mu_0)}\rho_B^-(\omega',
\mu_0)\,,
\end{eqnarray}
where
\begin{eqnarray}
&& V(\mu, \mu_0)  =-\int_{\alpha_s(\mu_0)}^{\alpha_s(\mu)}{d
\alpha\over \beta(\alpha)}\bigg[\Gamma_{\rm
cusp}(\alpha)\int_{\alpha_s(\mu_0)}^\alpha{d\alpha'
\over \beta(\alpha')}+\gamma_+(\alpha)\bigg],\nonumber \\
&& g(\mu, \mu_0)  =-\int_{\alpha_s(\mu_0)}^{\alpha_s(\mu)}d
\alpha{\Gamma_{\rm cusp}(\alpha)\over \beta(\alpha)}.
\end{eqnarray}
Using the orthogonality of the Bessel function, we can express
$\phi_B^-(\omega,\mu)$ in terms of $\rho_B^-(\omega', \mu)$:
\begin{eqnarray}
\phi_B^-(\omega,\mu)=\int^\infty_0\frac{d\omega'}{\omega'}J_0(2\sqrt{\frac{\omega}{\omega'}})\rho_B^-(\omega',\mu).\label{inversetrans}
\end{eqnarray}

Substituting Eq.(\ref{inversetrans}) into Eq.(\ref{factorization
of correlator:tree}), we obtain the factorization formula of
the correlation function in the dual space:
\begin{eqnarray}
\Pi_{\rm T}(\mu, n\cdot p) &=& \tilde{f}_B(\mu) \, m_B \,
C_T^{(-)}(n \cdot p, \mu) \int_0^{\infty} {d \omega' \over
\omega'}~ j_T^{(-)}(\hat{\omega}^{\prime},\mu) \,
\rho_B^{-}(\omega',\mu) \,\nonumber \\
&+& \tilde{f}_B(\mu) \, m_B  C_T^{(+)}(n \cdot p, \mu)
\int_0^{\infty} {d \omega \over \omega-\bar n\cdot p}~
J_T^{(+)}({\omega},\mu) \, \phi_B^{+}(\omega,\mu) , \label{NLO
factorization formula of correlator2 }
\end{eqnarray}
where $j_T^{(-)}(\hat{\omega}^{\prime},\mu)$ is the jet function
in the dual space. In the second term of the RHS,
  $J_T^{(+)}(\omega)$ and $\phi_B^+(\omega)$ are not transformed to the dual space because
  it is not necessary to perform the RG evolution to this term,
 which will be explained in the first comment in the next subsection.
  We rewrite $j^{-}(\hat{\omega}^{\prime},\mu)$
  as $j(\hat{\omega}^{\prime},\mu)$ which satisfies the following RG equation
\begin{eqnarray}
&& {d \over d \ln \mu} j(\hat{\omega}^{\prime},\mu)   =
\Gamma_j(\mu)j(\hat{\omega}^{\prime},\mu) \,. \label{jetrge}
\end{eqnarray} The factorization-scale independence of the correlation function indicates
\begin{eqnarray}
&& \Gamma_j(\mu)  =
-\Gamma_C(\mu)-\Gamma_\rho(\mu)-\tilde{\gamma}(\alpha_s(\mu)),\label{gammarealtion}
\end{eqnarray}
where $\tilde{\gamma}(\alpha_s)$ is the anomalous dimension of $\tilde{f}_{B}(\mu)$. The RG equation of
$\tilde{f}_B(\mu)$ is
\begin{eqnarray}
{d \over d \ln \mu} \, \tilde{f}_B(\mu)
=\tilde{\gamma}(\alpha_s)\, \tilde{f}_B(\mu) \,,
\end{eqnarray}
where
\begin{eqnarray}
\tilde{\gamma}(\alpha_s) &=& {\alpha_s \, C_F \over 4 \pi} \,
\left [ \tilde{\gamma}^{(0)} +
\left ({\alpha_s \over 4 \pi} \right )\, \tilde{\gamma}^{(1)} + ...  \right ] \,, \nonumber \\
\tilde{\gamma}^{(0)} &=& 3 \,, \qquad \tilde{\gamma}^{(1)}= {127
\over 6} + {14\, \pi^2 \over 9} - {5 \over 3}\, n_f
\,,\label{eq:tlfb}
\end{eqnarray}
and $n_f$ is the number of light quark flavors. The evolution factor of $\tilde{f}_{B}(\mu)$ is:
\begin{eqnarray}
U_2(\mu_{h2},\mu ) &=& {\rm Exp}  \bigg [
\int_{\alpha_s(\mu_{h2})}^{\alpha_s(\mu)} \,
d \alpha_s \, \frac{\tilde{\gamma}(\alpha_s)}{\beta(\alpha_s)} \bigg ] \, \nonumber \\
&=& z^{- \frac{\tilde{\gamma}^{(0)} }{2 \,\beta_0} \, C_F} \bigg [1+
\frac{\alpha_s(\mu_{h2}) \, C_F}{4 \pi}  \, \left (
{\tilde{\gamma}^{(1)} \over 2 \, \beta_0} - {\tilde{\gamma}^{(0)}
\, \beta_1 \over 2 \, \beta_0^2 } \right ) (1-z) +{\cal
O}(\alpha_s^2) \bigg ]\,,
\end{eqnarray}
where $z=\alpha_s(\mu)/\alpha_s(\mu_{h2})$.

The anomalous dimension of the jet function can be expressed as
\begin{eqnarray}
&& \Gamma_j  = \Gamma_{\rm cusp}(\alpha_s) \ln { \mu^2 \over n
\cdot p\, \hat {\omega}'}+\gamma_{hc}(\alpha_s),
\end{eqnarray}
where ${\gamma}_{hc}(\alpha_s) = {\alpha_s \, C_F \over 4 \pi} \, [
{\gamma}_{hc}^{(0)} + ({\alpha_s \over 4 \pi} )\,
{\gamma}_{hc}^{(1)} + ...  ] $ . At one-loop level,
$\gamma^{(0)}_{hc}=0$. There is no calculation about the
two-loop anomalous dimension $\gamma^{(1)}_{hc}$ till now. This
parameter could not be determined by the
factorization-scale independence of the correlation function
($\gamma_{hc}^{(1)}=-\gamma_h^{(1)}+\gamma^{(1)}_+-\tilde{\gamma}^{(1)}$), since $\gamma^{(1)}_+$ is unknown yet.  However, it
has been checked numerically that the  form factors are
insensitive to $\gamma^{(1)}_{hc}$. In addition, NLO corrections should not be very large for the convergence of the 
$\alpha_{s}$ expansion.
We choose $\gamma^{(1)}_{hc}=0$ in our calculation. The
solution of Eq.(\ref{jetrge}) can be obtained straightforwardly
\begin{eqnarray}
&&  j(\hat{\omega}^{\prime},\mu) =
e^{-2V_{hc}(\mu,\mu_{hc})}\left(\frac{\mu_{hc}^2}{\hat{\omega}'\bar{n}\cdot
p}\right)^{g(\mu,\mu_{hc})} j(\hat{\omega}^{\prime},\mu_{hc}) \,,
\end{eqnarray}
with $j(\hat {\omega}',\mu_{hc})$ given by
\begin{eqnarray}
 j(\hat{\omega}^{\prime},\mu_{hc}) &&
=2K_0\left(2\sqrt{\frac{1}{\eta'}}\right)\bigg\{1 + \frac{\alpha_s
\, C_F}{4 \, \pi} \, \bigg [ \ln^2 { \mu^2_{hc} \over -p^2 }
-{\pi^2 \over 3} -1 - \frac{1}{2}\ln \hat{\eta}'(2 \ln {
\mu^2_{hc} \over -p^2 }+3)\nonumber \\ &&+\frac{1}{4}\ln^2
\hat{\eta}' \bigg ]\bigg\}+\frac{\alpha_s \, C_F}{ \, 2\pi}
K^{(2,0)}_0 \left(2\sqrt{\frac{1}{\eta'}} \right ) +\frac{\alpha_s
\, C_F}{
\pi}\int_{2\sqrt{1\over\eta'}}^\infty\frac{d\beta}{\beta}K_0(\beta)
\,\,\,\, ,\label{dualjet}
\end{eqnarray}
where $\hat{\eta}'=e^{-2\gamma_E}
\eta'=-\hat{\omega}'/\bar{n}\cdot p$. The detailed derivation of
$j(\hat {\omega}',\mu_{hc})$ is given in the Appendix A.
Collecting evolution factors of the hard function, the jet
function, the $B$-meson LCDA and the $B$-meson decay constant
together, we obtain the RG improved correlation function
\begin{eqnarray}
\Pi_{\rm T}(\bar{n}\cdot p) &=& m_B    \, \Big [U_1(n \cdot
p,\mu_{h1},\mu ) \, U_2(\mu_{h2},\mu ) \Big ] \, \Big [
\tilde{f}_B(\mu_{h2}) \,
{C}^{(-)}_{T}(n \cdot p, \mu_{h1})  \Big ] ~\nonumber \\
&& \times \int_0^{\infty} {d \omega' \over \omega'}~
\left[U_j(n \cdot p,\mu_{hc},\mu ) \, U_\rho(n \cdot p,\mu_{0},\mu ) \right]\, j(\hat{\omega}^{\prime},\mu_{hc}) \, \rho_B^{-}(\omega',\mu_0) \, \, \nonumber \\
&& + \, m_B   \left [U_1(n \cdot
p,\mu_{h1},\mu ) \, U_2(\mu_{h2},\mu ) \right ] \, \tilde{f}_B(\mu_{h2}) \,
  \int_0^{\infty} d \omega \,{ \phi_B^{+}(\omega,\mu) \over \omega-
\bar n \cdot p}~ J^{(+)}_{T}\left({\omega \over \bar n \cdot p}\right)
\,  \,, \label{resummation improved
factorization formula}
\end{eqnarray}
where
\begin{eqnarray}
U_j(n \cdot p,\mu_{hc},\mu )&=&
e^{-2V_{hc}(\mu,\mu_{hc})}\left(\frac{\mu_{hc}^2}{\hat{\omega}'\bar{n}\cdot
p}\right)^{g(\mu,\mu_{hc})} \,,\nonumber \\
 U_\rho(n \cdot p,\mu_{0},\mu )  &=&
e^{V(\mu,\mu_0)}\left(\frac{\mu_0}{\hat{\omega}'}\right)^{-g(\mu,\mu_0)}\,.
\end{eqnarray}

We are now in the position to construct the sum rules for the three
$B \to \pi$ form factors at NLO. Useful dispersion integrals
are collected in the Appendix B. Performing the Borel transformation,
we obtain the sum rules for the $B\to \pi$ form factors. The
tensor form factor reads
\begin{align}
f_\pi e^{-m_\pi^2n\cdot p/\omega_M^2}f_{B\pi}^T(q^2)= 
 \left[U_1(n \cdot p,\mu_{h1},\mu ) U_2(\mu_{h2},\mu ) \right]  \,
\tilde{f}_B(\mu_{h2}) &
\nonumber \\
 \times  \int_0^{\omega_s} {d
\omega}~e^{-\omega/\omega_M} \bigg[{C}_T^{(-)}(n \cdot p,
\mu_{h1})\,\rho^-_{eff}(\omega)-\phi^+_{B,eff}(\omega) \bigg] & , \label{sumrule1}
\end{align}
where
\begin{eqnarray}
\phi_{B, \rm eff}^{+}(\Omega, \mu) =&&  \frac{\alpha_s \, C_F}{4
\, \pi} \,\, \int_{\Omega}^{\infty} \,\, {d \omega \over \omega}
\,\, \phi_{B}^{+}(\omega, \mu) \,\,\,,
\label{effective B-meson plus}\nonumber  \\
 \rho^-_{eff}(\Omega, \mu)=&& \int_0^{\infty} {d \omega' \over \omega'}
\bigg\{\bigg[1 + \frac{\alpha_s \, C_F}{4 \, \pi} \, \bigg ( \ln^2
{ \mu^2 \over n\cdot p\Omega }-2\ln { \mu^2 \over n\cdot p\Omega
}\ln{\hat\omega'\over
\Omega}\nonumber \\
&&+{1\over 2}\ln^2{\hat\omega'\over \Omega} -{3\over
2}\ln{\hat\omega'\over \Omega}+{\pi^2\over 2} -1\bigg )\bigg]J_0
\left({2\sqrt{\frac{\Omega}{\omega'}}}\right)\nonumber \\
&&+\frac{\alpha_s \, C_F}{4 \, \pi} \left(\ln {\hat\omega'\over
\Omega}+{3\over 2}\right)\pi N_0
\left({2\sqrt{\frac{\Omega}{\omega'}}}\right)\nonumber \\
&&+\frac{\alpha_s \, C_F}{ \, 2\pi}\bigg [J^{(2,0)}_0
\left({2\sqrt{\frac{\Omega}{\omega'}}}\right)
+\frac{\Omega}{\omega'}
{_2F_3}(1,1;2,2,2;-\frac{\Omega}{\omega'})-\ln\frac{\Omega}{\hat{\omega}'}\bigg]\bigg\}\nonumber \\
&&\times U_j(\mu_{hc},\mu ) U_\rho(\mu_{0},\mu
)\rho_B^{(-)}(\omega',\mu_0) \,,\label{sumrule2}
\end{eqnarray}
with $_pF_q(a_1...a_p;b_1...b_q;z)$ being the generalized
hypergeometric function. Vector and scalar form factors
$f^{+,0}_{B\pi}$ have similar forms
\begin{align}
 f_{\pi} \,\, e^{-m_{\pi}^2/(n \cdot p \, \omega_M)} 
\left \{ r^{\prime} \, f_{B \pi}^{+}(q^2)
\,, \,  f_{B \pi}^{0}(q^2)  \right \}    = \left[ U_1(n \cdot p,\mu_{h1},\mu ) U_2(\mu_{h2},\mu )\right ] \, \tilde{f}_B(\mu_{h2}) 
\int_0^{\omega_s} \,\, d \omega^{\prime}  
\, e^{-\omega^{\prime} / \omega_M} & 
\,\,\,   \nonumber \\
\times   \, \bigg \{ \left [
{C}^{(-)}_{\bar{n}} -\overline{r^{\prime}} 
C^{(-)}_{n} 
\right ]  (n \cdot p, \mu_{h1})
\, \rho_{B, \rm eff}^{-}(\omega^{\prime}, \mu)  
+\left(r\mp\overline{r^{\prime}}\right)\, \phi_{B, \rm eff}^{+}(\omega^{\prime}, \mu)
 \bigg
\} \, & ,\label{NLO sum rules of form factors}
\end{align}
where $r^{\prime}=n\cdot p/m_{B}$. For above results, several comments are as follows:
\begin{itemize}
\item We perform the complete RG evolution of $\phi_{B}^{-}$ terms. For $\phi_{B}^{+}$ terms, we only apply the RG evolution to the hard function and the $B$-meson decay constant not to $J^{(+)}_{i}$ and $\phi^{+}_{B}$ since firstly the anomalous dimension of $J^{(+)}_{i}$ is unknown yet and secondly evolution effects of the jet function and the $B$-meson LCDA will partially cancel each other. In principle we should resum logarithmic terms in $J^{(+)}_{i}$ and $\phi^{+}_{B}$ to leading-logarithm level because the jet function starts at $\mathcal{O}(\alpha_{s})$. While due to the factorization-scale independence of the correlation function, we set $\mu$ to be a hard-collinear scale in the numerical analysis, which means that there are no large logarithms in $J^{(+)}_{i}$ and $\phi^{+}_{B}$. Also uncertainties, arising from the variation of $\mu$, of these $\phi^{+}_{B}$ terms should be small since these terms are suppressed by $\alpha_{s}$. As we can see from the numerical analysis, the factorization-scale dependence of these terms is negligible.
\item {Dispersion integrals of the correlation function in the
 momentum space is non-trivial for the appearance of both pole
 and branch-cut singularities, which can be seen in the
 appendix of \cite{ymwang:2015wf}.
 While in the dual space, the imaginary part of the jet function in Eq.(\ref{sumrule2})
 can be obtained much more easily, i.e., we can simply make the replacement $\bar{n}\cdot p\to \Omega e^{i\pi}$}.
\end{itemize}

\section{NUMERICAL ANALYSIS}
\label{section: numerical}

In this section, we perform the numerical analysis of the
$B\to \pi$ form factors.
$B$-meson LCDAs serve
as fundamental ingredients of the $B$-meson LCSR approach. Nevertheless there is a
very limited knowledge about these LCDAs so far. Several
phenomenological models of the $B$-meson LCDA $\phi^{+}_{B}$ are suggested.
We employ three typical
models \cite{Grozin:1996pq,Braun:2003wx,Kawamura:2001jm}:
\begin{eqnarray}
 &&  \phi_{B1}^+(\omega,\mu_0) =
\frac{\omega}{\omega_0^2} \, e^{-\omega/\omega_0} \,,
\nonumber \\
&&  \phi_{B2}^+(\omega,\mu_0)= \frac{\omega}{2 \,\omega_0} \,
\theta(2\omega_0-\omega),
\nonumber \\
&&  \phi_{B3}^+(\omega,\mu_0)= \frac{3 }{4\omega_1^3 } \,
 \theta(2\omega_1-\omega)\omega(2\omega_1-\omega) \,,\,
\omega_1={3\over 2}\omega_0.
 \label{models of B-meson LCDAs}
\end{eqnarray}
Neglecting the contribution from the three-particle Fock state, $\phi_{B}^-(\omega,\mu_0)$ is determined by the Wandzura-Wilczek approximation
\cite{Beneke:2000wa}
\begin{eqnarray}
\phi_{B}^-(\omega,\mu_0) = \int_0^1 \, { d \xi \over \xi} \,
\phi_{B}^+ \left ({\omega \over \xi}\,, \mu_0 \right ) \,.
\label{WW relation}
\end{eqnarray}
The parameter $\omega_0$ equals the inverse moment of
$B$-meson LCDAs, i.e.,
\begin{eqnarray}
\, \lambda_B^{-1}(\mu) &=& \int_0^{\infty} \, {d \omega \over
\omega} \, \,\phi_{B}^+(\omega,\mu) = \frac{1}{\omega_{0}}\,.
\end{eqnarray}
$\lambda_{B}$ is closely related to exclusive $B$-meson decays.
The experimental data can only
give a very rough constraint on this parameter. In the previous work
\cite{ymwang:2015wf}, the inverse moment is determined by
fixing $f_{B\pi}^{+} (0)=0.28 \pm 0.03$
which is the prediction of the pion LCSR
\cite{Khodjamirian:2011ub}. In this paper we adopt the same
value as in \cite{ymwang:2015wf} for comparison, i.e.,
$\omega_0=0.354^{+0.038}_{-0.030}$(GeV).

\begin{figure}[!ht]
\begin{center}
\includegraphics[width=0.4 \columnwidth]{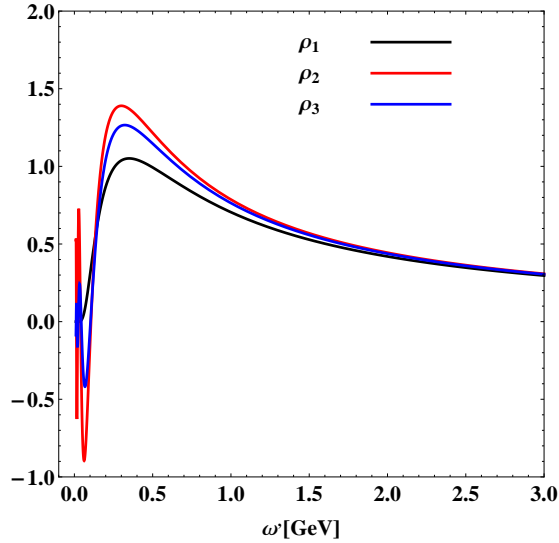}\\
\caption{Shapes of $B$-meson LCDAs in the dual space.
Black, red and blue lines correspond to
$\rho^{-}_{B1}{(\omega',\mu)}$, $\rho^{-}_{B2}{(\omega',\mu)}$ and $\rho^{-}_{B3}{(\omega',\mu)}$,
respectively.}
 \label{fig:dualwave}
 \end{center}
\end{figure}

Because the RG equation of $\phi^{-}_{B}$ is evaluated in the dual
space, the corresponding dual-space expression of this LCDA is
required.  $\rho^{-}_{B}$ has a similar meaning with
Gegenbauer moments
 of the light-meson LCDAs. Explicit forms of $\rho^{-}_{B}$ are
\begin{eqnarray}
\rho_{B1}^{-}(\omega',\mu)=\rho_{B1}^{+}(\omega',\mu)&=&{1\over\omega'}e^{-\omega_0/\omega'},
\nonumber
\\
\rho_{B2}^{-}(\omega',\mu)=\rho_{B2}^{+}(\omega',\mu)&=&\frac{1}{2\omega_0}J_2\left(2\sqrt{\frac{2\omega_0}{\omega'}}\right),
\nonumber
\\
\rho_{B3}^{-}(\omega',\mu)=\rho_{B3}^{+}(\omega',\mu)&=&\frac{3}{4\omega_1}\sqrt{\omega'\over
2\omega_1}J_3\left(2\sqrt{2\omega_1\over\omega' }\right).{\label
{dual wavefunction}}
\end{eqnarray}
To give a more intuitive picture of LCDAs in the dual space,
we plot the $\omega'$ dependence of them in Fig.\ref{fig:dualwave}. The dual-space LCDAs are factorization-scale
dependent. The RG evolution effect can modify the behavior of
the original model \cite{Feldmann:2014ika}, and this effect has been
considered in our calculation of the form factors.

\begin{figure}[!ht]
\begin{center}
\includegraphics[width=0.45 \columnwidth]{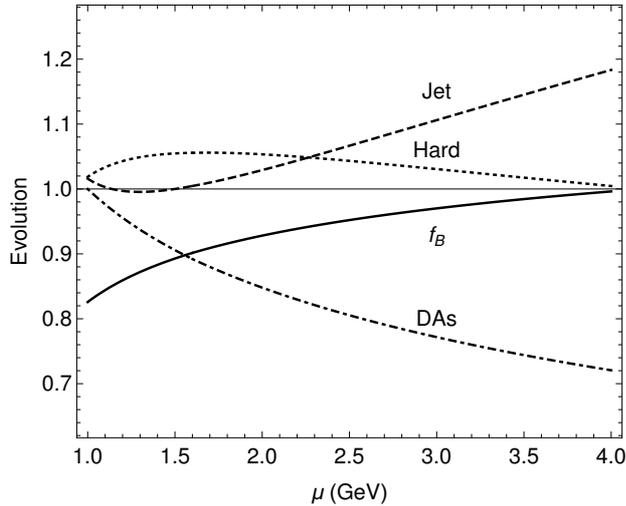}\\
 \caption{Evolution  factors of the hard function, the $B$-meson decay constant, the jet function and the $B$-meson LCDA,
 which are denoted using dotted, solid, dashed and dot-dashed lines, respectively.} \label{fig:Evolution}
\end{center}
\end{figure}

\begin{figure}[!ht]
\begin{center}
\includegraphics[width=0.45 \columnwidth]{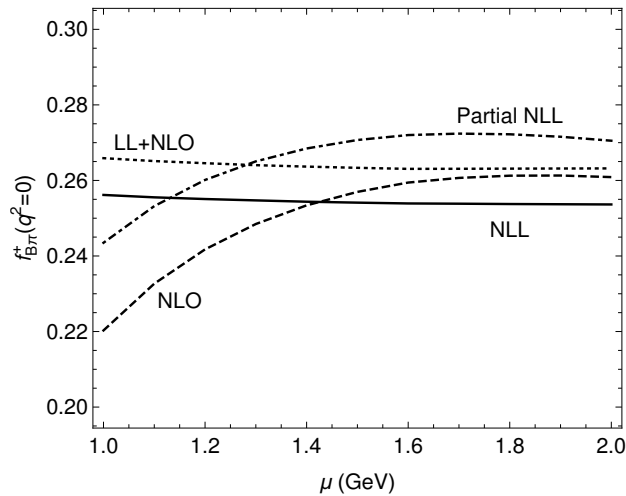}\\
 \caption{The factorization-scale dependence of $f^{+}_{B\pi}(0)$. Solid, dotted, dot-dashed and dashed lines
  stand for values of the form factor with full RG evolution at NLL level, at LL level, with RG evolution only respect to the hard coefficient and the $B$-meson decay constant and without RG evolution, respectively.
 } \label{fig:mudep}
\end{center}
\end{figure}

Before presenting numerical results of the form factors, we
first show behaviors of evolution factors of the hard
function ($U_1(\mu,\mu_{h1})$), the jet function
($U_j(\mu,\mu_{hc})$), the $B$-meson LCDA ($U_{}\rho(\mu,\mu_{0})$) and the
$B$-meson decay constant ($U_2(\mu,\mu_{h2})$), in
Fig.\ref{fig:Evolution}. $\omega'$ is fixed at $1.0$GeV when
plotting this figure. This choice leads to large logarithmic
terms in evolution kernels of the LCDA and the jet function,
hence evolution effects of these two functions are significant.
But there is a strong cancellation between these two effects due to different signs of slopes of their curves. This
cancellation is important to guarantee the scale invariance of the
form factors. To illustrate the effect of the RG evolution, we plot in Fig.\ref{fig:mudep} the scale dependence of
$f^{(+)}_{B\pi}(0)$, where the first type of $B$-meson LCDAs
$\phi^{\pm}_{B1}(\omega)$ and $\rho^{-}_{B1}(\omega')$ are
employed. It is obvious that after the complete RG evolution, the
theoretical prediction of the form factor is almost independent of
the factorization scale as expected. Results of
$f^{(+)}_{B\pi}(0)$ with leading logarithm (LL) resummation and
next-to-leading  logarithm (NLL) resummation are both displayed for a 
comparison. It can be seen that the scale dependence is mild in both
cases, but the NLL resummation reduces the value of the form factor
about $3\%$ compared to the LL-resummation value.  For terms contain $\phi_B^+$, the RG evolution
has not been performed (suppressed by the coupling constant). This figure also indicates that $\phi_{B}^{+}$ terms of the form factors are almost factorization-scale independent. In the numerical analysis, we set the factorization scale to be a hard-collinear scale ($\mu= 1.5 \pm0.5{\rm GeV}$),
as there are no large logarithmic terms in $\phi^{+}_{B}$ terms at this
scale \cite{ymwang:2015wf}.

\begin{figure}[!ht]
\begin{center}
\includegraphics[width=0.45 \columnwidth]{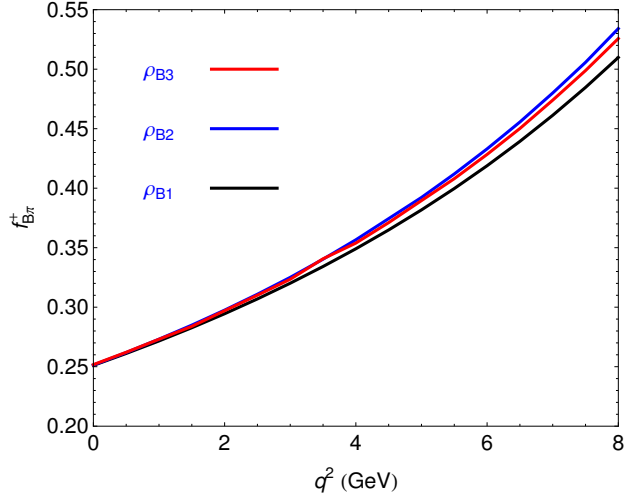}\\
\caption{The  LCDA-model dependence of the vector form factor.
Black, blue and red lines stand for the form factor
computed with $\rho^{-}_{B1}{(\omega',\mu)}$,
$\rho^{-}_{B2}{(\omega',\mu)}$ and
$\rho^{-}_{B3}{(\omega',\mu)}$, respectively.}
 \label{fig:wfdep}
 \end{center}
\end{figure}

Since $B$-meson LCDAs are most important inputs of the
$B$-meson LCSR, we need to test the LCDA-model dependence of the form
factors. In Fig.\ref{fig:wfdep}, the vector form factor computed
with three different B-meson LCDA models is displayed. Central values of the inverse moment are fitted as 0.392 in
$\rho^{-}_{B2}$ and 0.382 in $\rho^{-}_{B3}$. From this figure, we
can see that model of the $B$-meson LCDA has a tiny
influence on the shape of the form factor. Hereafter we will take
$\rho^{-}_{B1}$ as the default model.

\begin{figure}[!ht]
\begin{center}
\includegraphics[width=0.46 \columnwidth]{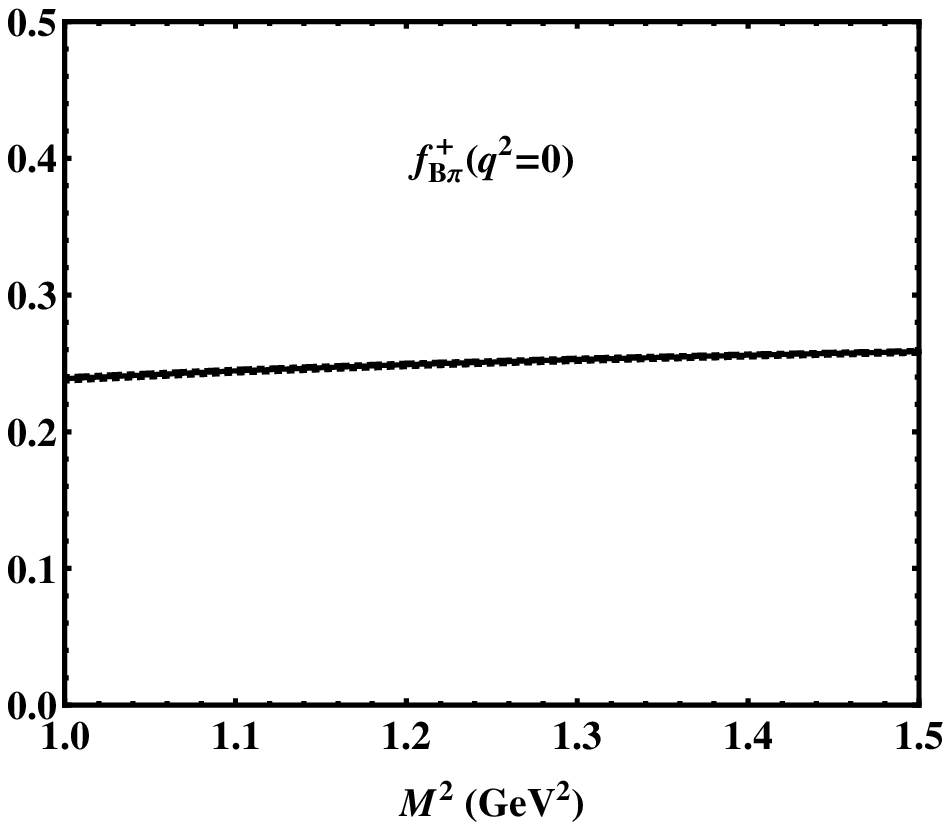}
\includegraphics[width=0.46 \columnwidth]{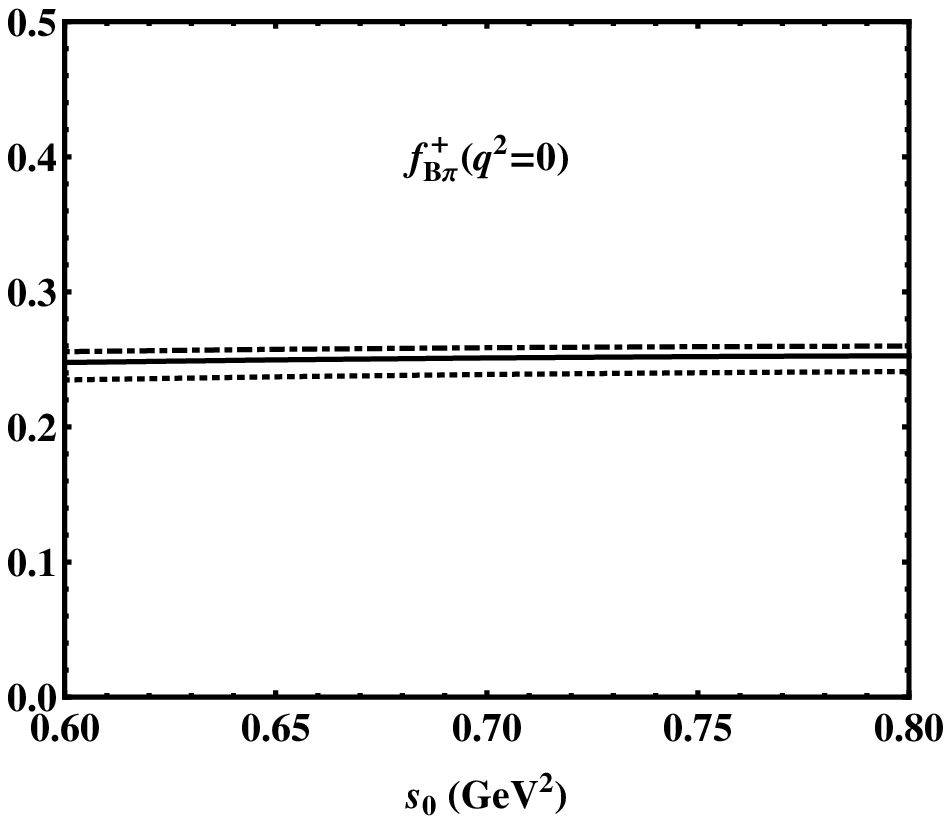}\\
\caption{The Borel parameter and the effective threshold  dependence of
$f^+_{B\pi}(0)$. Solid, dot-dashed and dotted lines in the left (right) figure correspond to $s_{0}=0.7{\rm GeV^2}$, $0.75{\rm GeV^2}$ and $0.65{\rm GeV^2}$ ( $M^{2}=1.25{\rm GeV^{2}}$, $1.0{\rm GeV^{2}}$ and $1.5{\rm GeV^{2}}$), respectively.}
 \label{fig:borel dependence}
\end{center}
\end{figure}

In the LCSR approach, the form factors should be insensitive to
the Borel parameter and the effective threshold. These parameters are constrained following conditions in \cite{ymwang:2015wf},
where the contribution from excited and continuum states should be less
than $40\%$ and the rate of change ${\partial \ln
f^T_{B\pi}(q^2)\over
\partial\ln \omega_M} \leq 35\%$. We fix $q^2=0$ to study $\omega_M$ and $s_0$
 dependence of the form factors. Above constraints lead to a region $0.24
\leq \omega_M \leq 0.36$ (corresponding to $1.0 \leq M^{2}/{\rm
GeV^{2}} \leq 1.5$) for all of the three form factors. We
plot the Borel mass dependence of the form factor $f^+_{B\pi}(0)$
in Fig.(\ref{fig:borel dependence}), a manifest platform at
$M^2\in [1.0 {\rm GeV}^2,1.5{\rm GeV}^2]$ guarantees that our
calculation is
 insensitive to this  unphysical parameter. The form factors are also
almost independent on the effective threshold $s_0$ when it is
adopted as $s_0=(0.7\pm0.05){\rm GeV^2}$.

It has been argued that the $B\to \pi$ form factors calculated
using  the $B$-meson LCSR can be trusted at $q^2 \leq q_{max}^2 =
8 \, {\rm GeV^2}$ (see \cite{Khodjamirian:2006st} for more
detailed discussions). To extrapolate the form factors calculated
with the $B$-meson LCSR at large recoil toward large momentum transfer, we
follow the same vein with \cite{ymwang:2015wf}, where the
$z$-series parameterization was employed. In this parameterization,
the cut $q^2$-plane (the branch cut is the $q^2>t_+$ region of the
real axis) is mapped onto the unit disk $|z(q^2, t_0)|<1$ via the
conformal transformation
\begin{eqnarray}
z(q^2, t_0) =
\frac{\sqrt{t_{+}-q^2}-\sqrt{t_{+}-t_0}}{\sqrt{t_{+}-q^2}+\sqrt{t_{+}-t_0}}\,,
\end{eqnarray}
where $t_{+}=(m_B + m_{\pi})^2$ denotes the threshold of continuum
states in the $B^{\ast}$-meson channel. The free parameter
$t_0 \in ( - \infty \,,  t_{+})$ determines the value of $q^2$
mapped onto the origin in the $z$ plane. One can  adjust the value
of $t_0$  to minimize the corresponding $z$ interval in the
region $q_{min}^2 \leq q^2 \leq q_{max}^2$, in order that the $z$-series expansion converges rapidly. Here we choose the same
value as that in \cite{Khodjamirian:2011ub}
\begin{eqnarray}
t_0 = t_{+}^2 - \sqrt{t_{+}-t_{-}} \, \sqrt{t_{+} -q_{min}^2}\,,
\end{eqnarray}
where $q_{min}^2= -6.0 \, {\rm GeV^2}$ and $t_{-} \equiv (m_B -
m_{\pi})^2$. Using the  $z$-series expansion and taking into
account the threshold $t_{+}$ behavior, one can obtain the
parametrization of each form factor.

Parametrizations of the vector and the scalar form factors  have been
given in \cite{Bourrely:2008za,Khodjamirian:2011ub}. The
parametrization of the tensor form factor is similar with that of
$f^+_{B\pi}(q^2)$ \cite{Li:2015cta}
\begin{eqnarray}
f^T_{B\pi}(q^2) = \frac{f^T_{B\pi}(0)}{1-q^2/m_{B^*}^2}
\Bigg\{1+\sum\limits_{k=1}^{N-1}b^T_k\,\Bigg(z(q^2,t_0)^k- z(0,t_0)^k \nonumber\\
-(-1)^{N-k}\frac{k}{N}\bigg[z(q^2,t_0)^N-
z(0,t_0)^N\bigg]\Bigg)\Bigg\} \label{z parametrization of
fplus}\,,
\end{eqnarray}
where the expansion coefficient(s) $b^{T}_k$ is (are) determined by
matching the large-recoil $f^T_{B\pi}(q^2)$ onto Eq.(\ref{z
parametrization of fplus}). As the interval in the $z$ plane is
constrained in a small region, it is reasonable to truncate the
$z$-series at $N=2$ in the practical calculation. Slop parameters $b_1, \, \tilde{b}_1, \, b_1^T$ are
collected in Table \ref{tab of fitted parameters}. Uncertainties from different sources, including the inverse
moment, the model of $B$-meson LCDA, the Borel parameter, the effective 
threshold, quark masses, et al, are taken into
account in our numerical analysis. In Table \ref{tab of
fitted parameters}, we collect parameters which arise
large uncertainties. 

\begin{table}[!ht]
\begin{center}
\begin{tabular}{|c|c|c|c|c|c|c|}
  \hline
  \hline
 Parameter & default & $\lambda_{B}(1\text{GeV})$  & $\omega_{M}$ & $\overline{m_{b}}(\overline{m_{b}})$ &
$s^{B}_{0}$ & LCDA \\
  \hline
  $f^+_{B\pi}(0)$ &$0.254$ & $^{+0.022}_{-0.024}$ & $^{+0.008}_{-0.012}$ &$^{+0.009}_{-0.011}$&
$^{+0.011}_{-0.006}$ & $^{+0.075}_{-0.000}$\\
\hline
  $f^T_{B\pi}(0)$ &$0.254$ & $^{+0.022}_{-0.024}$ & $^{+0.007}_{-0.012}$ &$^{+0.009}_{-0.012}$&
$^{+0.011}_{-0.006}$ & $^{+0.076}_{-0.000}$\\
 \hline
  $b_1$ & $-4.13$ & $^{+0.08}_{-0.09}$ & $^{+0.02}_{-0.05}$ & $^{+0.00}_{-0.01}$ &
$^{+0.00}_{-0.01}$ & $^{+0.64}_{-0.00}$  \\
\hline
  $\tilde{b}_1$ & $-5.56$ & $^{+0.11}_{-0.12}$ & $^{+0.03}_{-0.06}$ & $^{+0.00}_{-0.01}$ &
$^{+0.00}_{-0.00}$ & $^{+0.79}_{-0.00}$  \\
\hline
  $b^T_1$ & $-4.42$ & $^{+0.08}_{-0.11}$ & $^{+0.02}_{-0.10}$ & $^{+0.00}_{-0.01}$ &
$^{+0.00}_{-0.01}$ & $^{+0.71}_{-0.00}$  \\
   \hline
   \hline
\end{tabular}
\end{center}
\caption{$z$-parameter fitted values of $f^{+}_{B
\pi}(0)$, $f^{T}_{B \pi}(0)$, $b_1$, $\tilde{b}_1$ and $b^T_1$ ($f^{0}_{B \pi}(0)$ is not listed
here because $f^{0}_{B \pi}(0)=f^{+}_{B \pi}(0)$). The notation
``default"   means that all of the parameters are taken as
central values.} \label{tab of fitted parameters}
\end{table}

\begin{figure}[!ht]
\begin{center}
\includegraphics[width=0.46 \columnwidth]{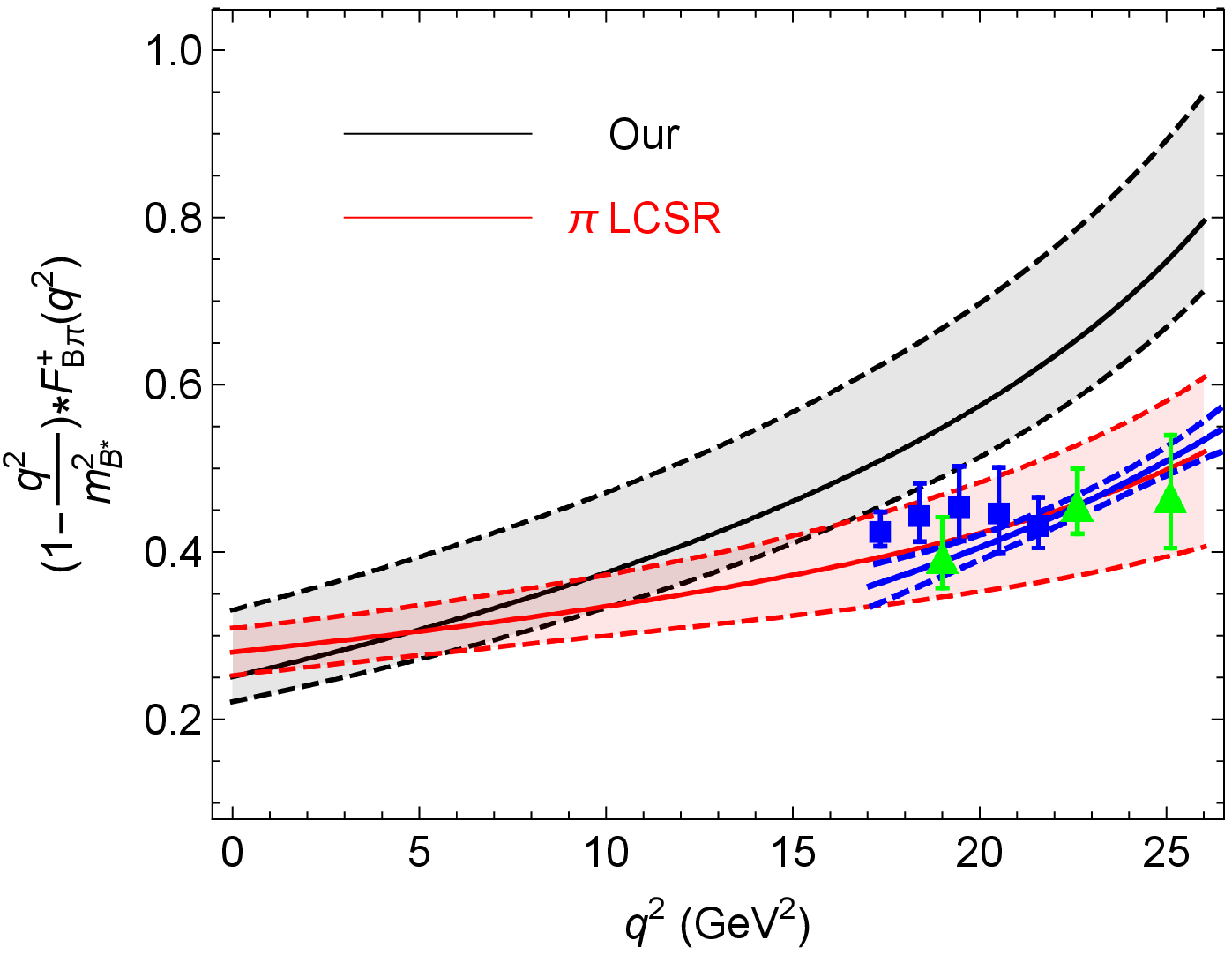}
\includegraphics[width=0.46 \columnwidth]{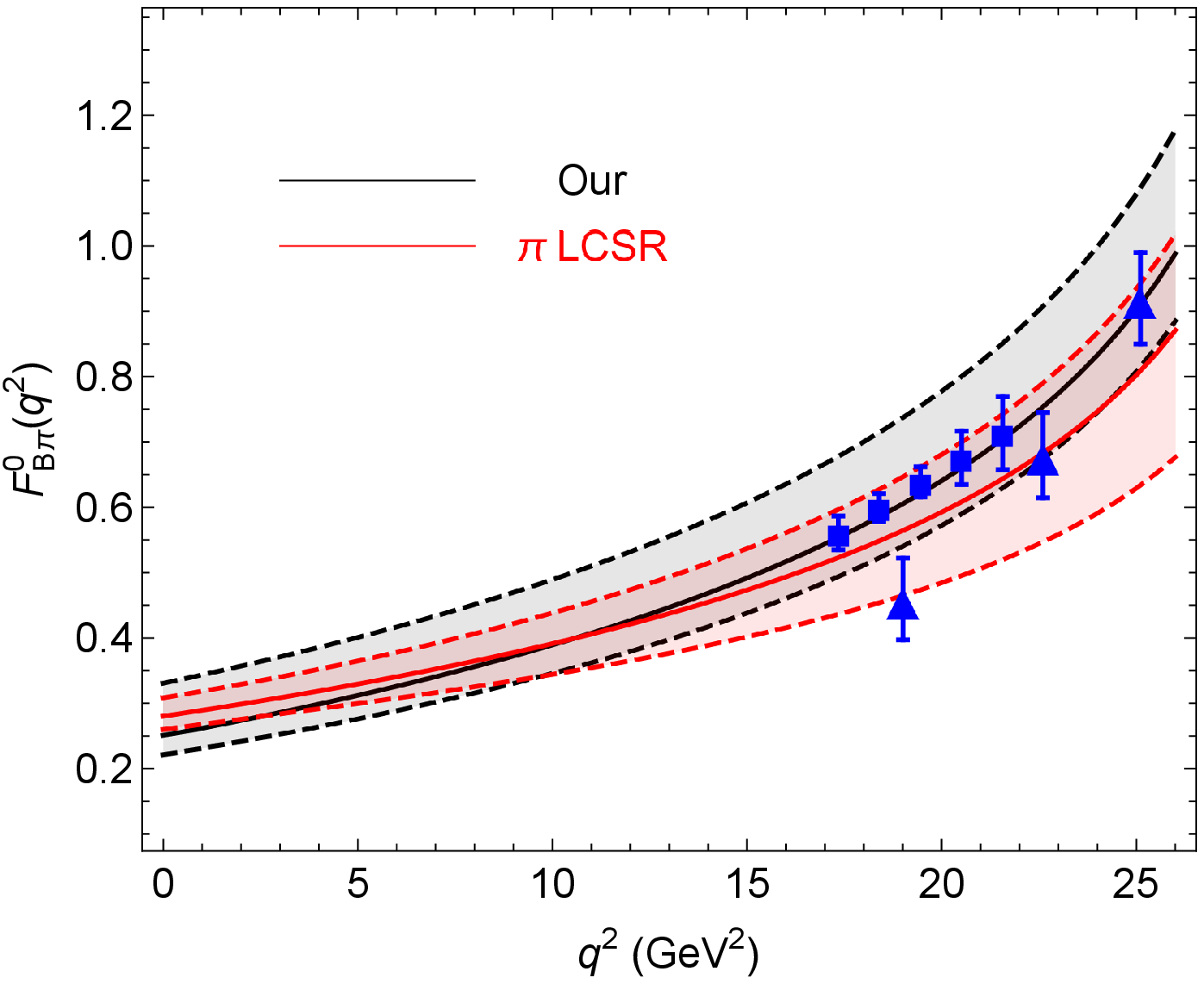}
\\ \qquad
 \caption{$q^2$ dependence of the form factor $f^{0}_{B \pi}(q^2)$,
and the re-scaled form factor $f^{+}_{B \pi}(q^2)$. Black and red curves are results of the $B$-meson and the pion LCSRs, respectively. Lattice QCD results are taken from HPQCD collaboration
\cite{Flynn:2015mha} (blue squres), Fermilab/MILC collaboration
\cite{Lattice:2015tia} (blue band) \label{fig:q2depvector} and RBC/UKQCD
collaboration \cite{Dalgic:2006dt} (green and blue triangles).}
\end{center}
\end{figure}

In Fig.(\ref{fig:q2depvector}) the $q^2$ dependence of form
factors $f^{0,+}_{B\pi}(q^2)$ are shown, where Lattice results are
from HPQCD collaboration \cite{Flynn:2015mha}, Fermilab/MILC
collaboration \cite{Lattice:2015tia} and RBC/UKQCD collaboration
\cite{Dalgic:2006dt}. Our results of $f^{0}_{B\pi}$ within errors
are in agreement with the Lattice data. While our results of
$f^{+}_{B\pi}$ is larger than the Lattice data. Results of $f^{0,+}_{B\pi}(q^2)$ from the pion
LCSR are also shown. It is manifest that slopes of these two form
factors with the $B$-meson LCSR are greater than that of the pion
LCSR. The difference between the $B$-meson and other approaches can
be understood through following points. 
(1) As we can see from Table 1, the parameter $\lambda_{B}$ brings huge uncertainty to results of the form factors. Values of the form factors at $q^{2}=0\,{\rm GeV^{2}}$ are significantly influenced by the changing of $\lambda_{B}$. But the value of this parameter is not determined yet.
(2) We only calculate leading-power contributions of the form factors in this work. While power-suppressed contributions, which are induced by higher-twist pion LCDAs, are taken into account in calculations of the pion LCSR. The subleading-power effect in the $B$-meson LCSR may influence both values of the form factors at $q^{2}=0\,{\rm GeV^{2}}$ and slopes of the form factors.

\begin{figure}[!ht]
\begin{center}
\includegraphics[width=0.7 \columnwidth]{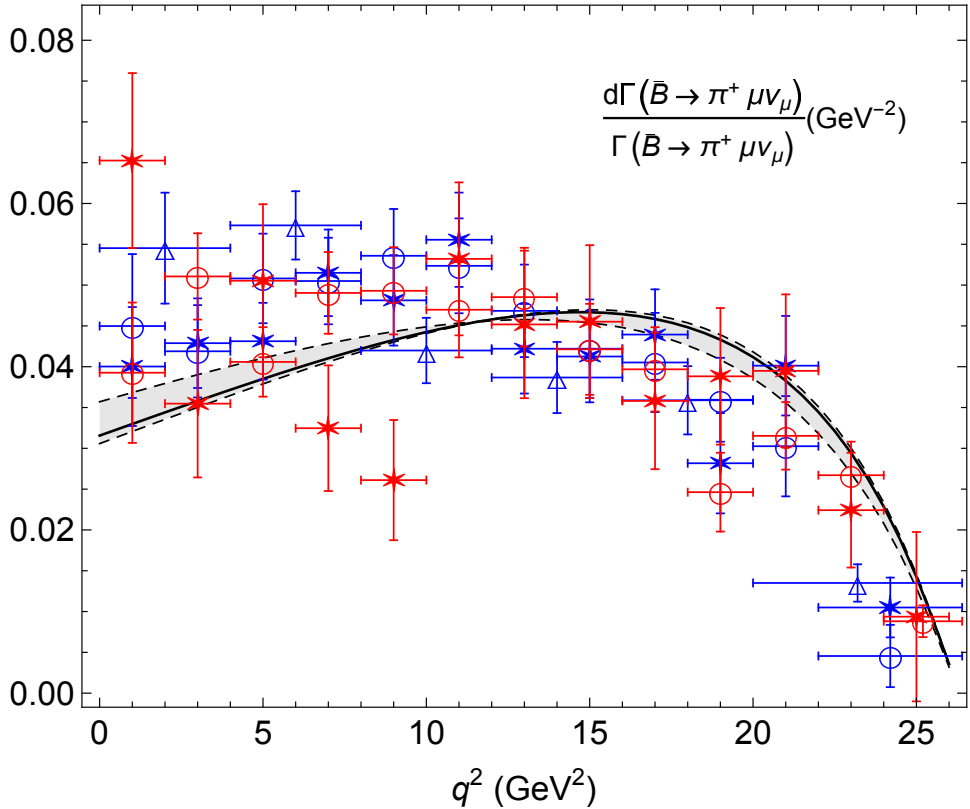}
\end{center}
 \caption{The normalized differential $q^{2}$ distribution of $B \to \pi \mu \nu_{\mu}$. The black solid curve represents the central value of our prediction and black dashed curves correspond to uncertainties. Experimental data bins are from \cite{Lees:2012vv} (blue stars), \cite{Sibidanov:2013rkk} (red stars), \cite{delAmoSanchez:2010af} (blue circles), \cite{delAmoSanchez:2010zd} (blue triangles) and \cite{Ha:2010rf} (red circles).}
\label{diffga.fkj}
\end{figure}

$B \to \pi$ form factors are very important
phenomenologically. Here we briefly discuss two applications of our
result. The CKM matrix element $|V_{ub}|$ can be determined from
the (partial) branching fraction of $B \to \pi \ell \nu_{\ell}$.
If we neglect mass of leptons, the integrated decay width is
written by
\begin{eqnarray}
\int_0^{ q_0^2} dq^2\frac{d \Gamma}{d q^2} \, (B \to \pi l \nu)
& \equiv
|V_{ub}|^2\Delta \zeta (0, q_0^2)\,, \label{differential
distribution formula: muon}
\end{eqnarray}
where $|\vec{p}_{\pi}|$ is the magnitude of the pion
three-momentum in the $B$-meson rest frame, $l=e,\mu$ and
\begin{eqnarray}
\Delta \zeta (0, q_0^2)=  \frac{G_F^2}{24 \pi^3} \,
\int_{0}^{q_0^2} \, d q^2 \, |\vec{p}_{\pi}|^3 \, \, |f_{B
\pi}^{+}(q^2)|^2  \,.
\end{eqnarray}
A straightforward extraction of $|V_{ub}|$ can be performed using
the relation
\begin{eqnarray}
|V_{ub}|^2=  {\Delta {\cal BR} (0, q_0^2)\tau_{B^0}\over\,\Delta
\zeta (0, q_0^2) \,},
\end{eqnarray}
where $\Delta {\cal BR} (0, q_0^2)$  is the integrated branching
ratio and the mean lifetime of the $B^0$ meson $\tau_{B^0} =(1.519 \pm
0.005) \, {\rm ps}$ \cite{Agashe:2014kda}.  Experimental
measurements of $\Delta {\cal BR} (0, q_0^2)$ of the semi-leptonic  $\bar
B^0 \to \pi^{+} \, \mu \, \nu_{\mu}$ decay
\cite{Lees:2012vv,Sibidanov:2013rkk} are given by
\begin{eqnarray}
\Delta {\cal BR} (0,  12 \, {\rm GeV^2}) &=& (0.83 \pm 0.03 \pm
0.04) \times 10^{-4} \,, \hspace{0.9 cm} {\rm [BaBar \,\, 2012]}
\nonumber \\
\Delta {\cal BR} (0,  12 \, {\rm GeV^2}) &=& (0.808 \pm 0.062)
\times 10^{-4} \,. \hspace{1.7 cm}  {\rm [Belle \,\,\,\,\,\,
2013]}
\end{eqnarray}
Utilizing  the result of the form factor $f_{B \pi}^{+}(q^2)$ which is
computed with the $B$-meson LCSR and extrapolated  with the $z$-series
parametrization we can obtain
\begin{eqnarray}
\Delta \zeta (0, 12 \, {\rm GeV^2}) &=& 4.93
\,\,{}^{+0.30}_{-0.05}\Big|_{\omega_0}
\,\,{}^{+0.36}_{-0.41}\Big|_{\omega_M}
\,\,{}^{+0.44}_{-0.23}\Big|_{s_0}
\,\,{}^{+2.79}_{-0}\Big|_{\phi_B}
\,\, \mbox{ps}^{-1}\nonumber\\
&=&4.93^{+2.99}_{-0.97}~\mbox{ps}^{-1}\,. \label{Delta Zeta 0 to
12}
\end{eqnarray}
Then the extracted CKM matrix element
\begin{eqnarray}
|V_{ub}|= \left(3.33^{+0.37}_{-0.74} |_{\rm th.} \pm 0.09 |_{\rm
exp.}\right)  \times 10^{-3} \,,
\end{eqnarray}
where the
theoretical uncertainty comes from uncertainties of $\Delta \zeta
(0, 12 \, {\rm GeV^2})$ as displayed in  (\ref{Delta Zeta 0 to
12}).
This $|V_{ub}|$ is larger compared to
\cite{ymwang:2015wf}, since the RG evolution
reduces the value of $f_{B\pi}^+(q^2)$. In Fig.\ref{diffga.fkj}, we display the normalized differential $q^{2}$ distribution of $B \to \pi \mu \nu_{\mu}$. Black curves represent the prediction of this work, where the solid line is the central value and dashed curves correspond to uncertainties. Due to the cancellation of the uncertainty of $f_{B\pi}^+(q^2)$, the uncertainty of the normalized differential distribution of $B \to \pi \mu \nu_{\mu}$ is small. Our prediction is in agreement with the experimental data from BarBar \cite{Lees:2012vv,delAmoSanchez:2010af,delAmoSanchez:2010zd} and Belle \cite{Sibidanov:2013rkk,Ha:2010rf}.

\section{CONCLUSION AND DISCUSSION}
\label{section: summaries}

We reviewed the method of calculating the $B \to \pi$ tensor
form factor with the $B$-meson LCSR. In this framework, the
method of regions was employed  and contributions from different
momentum regions are separated naturally. Precise soft
cancellation guarantees the factorization theorem. The
correlation function was factorized into the convolution of the hard
function, the jet function and the $B$-meson LCDA which correspond to contributions from 
hard, hard-collinear and soft regions, respectively. We obtained
one-loop-level  hard and  jet functions through the
analysis of symmetry-breaking effects.

To resum large logarithmic terms in  the form factors, we
carried out the complete RG evolution of the factorized
correlation function, including evolutions of the jet function and the $B$-meson LCDA. The
$B$-meson LCDA, defined via the HQET, obey the
Lange-Neubert equation which contains non-diagonal anomalous
dimension. Following the approach in \cite{Bell:2013tfa},
we diagonalized the RG equation of the $B$-meson LCDA in the dual
space and solved the diagonalized RG equation. The
same method was also applied to the evolution of the jet function.
Combining the evolution of each part together, we obtained the 
RG improved $B\to\pi$ form factors.

On the numerical side, we checked behaviors of the four evolution
kernels ($U_{1}, \, U_{2}, \, U_{j} \, {\rm and} \, U_{\rho}$) and illustrated cancellation effects among the kernels. We examined
the factorization-scale dependence of the RG improved form factors
and compared
our predictions with previous
results. We extrapolated the $q^2$ dependence of the form factors
to the whole physical region using the $z$-series expansion. Then we
compared values of the form factors in this work with that in the LQCD and
the pion LCSR. Phenomenologically we extracted the CKM matrix element $\vert V_{ub} \vert$ and analysed the normalized differential $q^{2}$ dependence of $B\to\pi\mu\nu_{\mu}$. The $B$-meson form factors
have many other phenomenological applications, such as the tensor form factor
can give important contributions to FCNC processes $B \to
(\pi,K)l^+l^-$. Of course a complete study of  phenomenological applications are
far more complicated, and we left it for the future work. This
work supplements the framework proposed in \cite{ymwang:2015wf},
and can be applied to various transition processes.

\section*{Acknowledgement}
We are grateful to Y. M. Wang for useful discussions and comments.
This work was supported in part by Natural Science Foundation of
Shandong Province, China under Grant No. ZR2015AQ006 and by
National Natural Science Foundation of China (Grants No. 11375208,
No. 11521505, No. 11235005, No. 11447009).

 \vspace{0.5 cm}

\appendix

\section{Jet function in the dual space}
\label{integrals}

The jet function in the dual space is defined by
\begin{eqnarray}
j^{(-)}\left({\mu_{hc}^2 \over n \cdot p \hat {\omega}'},{\hat
{\omega}' \over \bar n \cdot p}\right)
&=&\int^\infty_0\frac{d\omega}{\omega-\bar{n}\cdot
p}J_0(2\sqrt{\frac{\omega}{\omega'}}){J}^{(-)}\left({\mu^2 \over n
\cdot p \, \omega},{\omega \over \bar n \cdot p}\right)\nonumber \\
&=&\int^\infty_0\frac{d\eta}{1+\eta}J_0(2\sqrt{\frac{\eta}{\eta'}}){J}^{(-)}(\eta,\mu),
\end{eqnarray}
where
\begin{eqnarray}
J^{(-)}(\eta, \mu)&=&1 + \frac{\alpha_s \, C_F}{4 \, \pi} \, \bigg
[ \ln^2 { \mu^2 \over  -p^2 } - 2 \ln {(1+\eta) }\ln { \mu^2 \over
-p^2 }\nonumber \\ &+&\ln^2(1+\eta)- \frac{\eta-2}{\eta}\ln
{(1+\eta) } -{\pi^2 \over 6} -1 \bigg ] \,.
\end{eqnarray}
Using the formula
\begin{eqnarray}
\int^\infty_0 \frac{xdx}{(x^2+k^2)^{1-\lambda}
}J_0(ax)=\frac{1}{\Gamma(1-\lambda)}(\frac{2k}{a})^\lambda
K_\lambda(ka),\label{Bessleinte1}
\end{eqnarray}
which is valid for $\lambda<3/4$, and performing derivative with
respect to $\lambda$, and taking the limit $\lambda\to 0$, we can
get
\begin{eqnarray}
\int^\infty_0\frac{d\eta}{1+\eta}J_0\left(2\sqrt{\frac{\eta}{\eta'}}\right)&=&2K_0\left(2\sqrt{\frac{1}{\eta'}}\right),\nonumber \\
\int^\infty_0\frac{d\eta}{1+\eta}J_0\left(2\sqrt{\frac{\eta}{\eta'}}\right)\ln(1+\eta)
&=&(\ln \eta'-2\gamma_E)K_0\left(2\sqrt{\frac{1}{\eta'}}\right),\nonumber \\
\int^\infty_0\frac{d\eta}{1+\eta}J_0\left(2\sqrt{\frac{\eta}{\eta'}}\right)\ln^2(1+\eta)
&=&[\frac{1}{2}\ln^2 \eta'-2\gamma_E\ln
\eta'+2\gamma_E^2-2\psi'(1)]K_0\left(2\sqrt{\frac{1}{\eta'}}\right)\nonumber
\\&+&2 K^{(2,0)}_0
\left(2\sqrt{\frac{1}{\eta'}}\right).\label{appen1}
\end{eqnarray}
Another useful equation is
\begin{eqnarray}
\int^\infty_0\frac{dx}{x}J_0(bx)\ln(1+x^2)dx=2\int_b^\infty\frac{K_0(\beta)}{\beta}d\beta.\label{appen2}
\end{eqnarray}
Taking the advantage of Eqs. (\ref{appen1}) and (\ref{appen2}),
one can obtain Eq. (\ref{dualjet}).

\section{Dispersion integrals}

To obtain final expressions of the form factors, we need to
extrapolate $\bar n \cdot p$ to physical region. For the
consistency of our derivation, we must have
\begin{eqnarray}
\int^\infty_0\frac{d\omega}{\omega-\Omega-i\epsilon}J_0\left(2\sqrt{\frac{\omega}{\omega'}}\right)
&=&2K_0\left(-2i\sqrt{\frac{\Omega}{\omega'}}\right). \nonumber \\
\end{eqnarray}
The above equation indicates that the branch cut of the root and
logarithmic function be along negative real axis, and
$-\Omega-i\epsilon=\Omega e^{-i\pi}$.  The following equation can
be derived from Eq. (\ref{Bessleinte1})
\begin{eqnarray}
\int^\infty_0\frac{d\omega}{\omega-\Omega-i\epsilon}(1-{\omega\over
\Omega})^\lambda J_0\left(2\sqrt{\frac{\omega}{\omega'}}\right)
&=&i\pi e^{i\lambda\pi}\left({\omega'\over
\Omega}\right)^{\lambda/2}
\frac{1}{\Gamma(1-\lambda)}H_\lambda^{(1)}\left(2\sqrt{\frac{\Omega}{\omega'}}\right).
\end{eqnarray}
From which we obtain following useful results:
\begin{eqnarray}
\int^\infty_0\frac{d\omega}{\omega-\Omega-i\epsilon}\ln(1-{\omega\over
\Omega}) J_0\left(2\sqrt{\frac{\omega}{\omega'}}\right)
&=&-{\pi^2\over
2}J_0\left(2\sqrt{\frac{\Omega}{\omega'}}\right)-{\pi\over
2}\ln{\hat\omega'\over
\Omega}N_0\left(2\sqrt{\frac{\Omega}{\omega'}}\right)\nonumber \\
&-&i\left[{\pi^2\over
2}N_0\left(2\sqrt{\frac{\Omega}{\omega'}}\right)-{\pi\over
2}\ln{\hat\omega'\over
\Omega}J_0\left(2\sqrt{\frac{\Omega}{\omega'}}\right)\right],\\
\int^\infty_0\frac{d\omega}{\omega-\Omega-i\epsilon}\ln^2(1-{\omega\over
\Omega}) J_0\left(2\sqrt{\frac{\omega}{\omega'}}\right)
&=&{i\pi\over 2}\left({1\over 2} \ln^2{\hat\omega'\over
\Omega}+i\pi\ln{\hat\omega'\over \Omega}-{\pi^2\over
3}\right)H^{(1)}_0\left(2\sqrt{\frac{\Omega}{\omega'}}\right)\nonumber \\
&+&i\pi J^{(2,0)}_0\left(2\sqrt{\frac{\Omega}{\omega'}}\right)-\pi
N^{(2,0)}_0\left(2\sqrt{\frac{\Omega}{\omega'}}\right).
\end{eqnarray}
Following a similar way, another useful result is also obtained
\begin{eqnarray}
\int^\infty_0\frac{d\omega}{\omega}\ln(1-{\omega\over \Omega})
J_0\left(2\sqrt{\frac{\omega}{\omega'}}\right)
&=&2i\pi\int^\infty_{2\sqrt{\frac{\Omega}{\omega'}}}\frac{d\beta}{\beta}H^{(1)}_0(\beta).
\end{eqnarray}
Taking the imaginary part of the above equation, we have
\begin{equation}
{\rm Im}\int^\infty_0\frac{d\omega}{\omega}\ln(1-{\omega\over
\Omega}) J_0\left(2\sqrt{\frac{\omega}{\omega'}}\right)
=2\pi\left[-\gamma_E+\frac{\Omega}{\omega'}
{_2F_3}(1,1;2,2,2;-\frac{\Omega}{\omega'})-\ln\frac{\Omega}{\hat{\omega}'}\right].
\end{equation}
 Having
all of above equations in hand, we get final results  in Eqs.
(\ref{sumrule1}) and (\ref{sumrule2}).

\end{document}